\documentclass[10pt,letterpaper,twocolumn]{article}
\usepackage{hthtml}
\usepackage{amsmath}
\usepackage{wrapfig}
\usepackage{bardiag}
\usepackage{graphicx}
\usepackage{times}
\usepackage{xspace}
\usepackage{fancyvrb}
\usepackage[margin=1.9cm,bottom=1cm]{geometry}
\newgray{green}{0.8}
\newgray{blue}{0.7}
\newgray{red}{0.6}
\usepackage{isolatin1}

\newcommand{\WebCheck}{\emph{WebCheck}\xspace} 
\newcommand{\WebTest}{\emph{WebTest}\xspace} 
\newcommand{\WebWrite}{\emph{WebWrite}\xspace} 
\newcommand{\WebTeach}{\emph{WebTeach}\xspace} 
\let\xxx=\LaTeX
\renewcommand{\LaTeX}{\xxx\xspace}
 
\title{\WebTeach in practice: the entrance test to the Engineering
faculty in Florence}
\date{}

\author{
	\begin{minipage}{2\columnwidth}
	\centerline{
	Franco Bagnoli$^{(1,5,6)}$, 
	Fabio Franci$^{(2,5)}$, 
	Francesco Mugelli$^{(3)}$, 
	Andrea Sterbini$^{(4)}$}\vspace{0.5cm}
	\small
	\begin{center}
	\begin{minipage}{0.7\columnwidth}
	\small
	\raggedright
	1) Dip.\ Energetica, Univ. Firenze, Firenze, Italy.
	\texttt{franco.bagnoli@unifi.it}\\
	2)  Dip.\  Sistemi e Informatica, Univ. Firenze, Firenze, Italy.
    \texttt{fabio@dma.unifi.it}\\		
	3)  Dip.\ Matematica Applicata, Univ. Firenze, Firenze, Italy.
    \texttt{mugelli@dma.unifi.it}\\
	4) Dip.\ Informatica, Univ. ``La
  Sapienza'', Roma, Italy. \texttt{sterbini@di.uniroma1.it}\\
	5) INFM and CSDC Firenze, Italy. $\qquad$ 6) INFN, Firenze, Italy
	\end{minipage}\\[.3cm]
	\begin{minipage}{0.9\columnwidth}
We present the \WebTeach project, 
formed by a web interface to database
for test management, a wiki site for the diffusion of teaching
material and student forums, and a suite for the generation of
multiple-choice mathematical quiz with automatic elaboration of forms.
This system has been massively tested for the entrance test to the
Engineering Faculty of the University of Florence, Italy. 
\\[0.2cm]
\end{minipage}
\end{center}
	\end{minipage}
	}

\begin{document}
\maketitle

\section{The \WebTeach project}
The \WebTeach~\cite{WebTeach} project started in 1999 as a simple web interface to a 
database for the 
management of  students scheduling for examinations. At present it has
grown to include three different tools: \WebCheck, which corresponds
to the original database interface, \WebWrite, a powerful version of the  
WikiWikiWeb~\cite{wiki} concept, and 
\WebTest, a set of tools for the generation and automatic evaluation of 
multiple-choice mathematical quizzes. 

All this work was done in a semi-volunteer way, pushed by the
appreciation of our colleagues and of students. Our system is gaining
consensus mainly because we try to implement all 
suggested enhancements. We also tried to delegate as much administrative and
management work as possible to users, both teachers and students. So
teachers have the freedom of adding 
 users, courses and examinations to
the database, and students have the power of editing most of pages in
the \WebWrite space.

As a comparison, 
 the official web-based service of the
university of Florence is systematically affected by the so called 
``system-manager bottleneck syndrome'', for which each
administrative change has to wait for official approval. 

After five years of improvements, \WebWrite was massively tested this
year in correspondence to the entrance test to the engineering
faculty of the University of Florence, 
 900 participants
for two mathematical tests. The system was used
for test management, immediate self-evaluation of answers, 
optical reading of the elaborates using inexpensive hardware, 
and a HOWTO forum to students. 

All the software was developed in \texttt{Perl} and \texttt{C} using
free tools on a Linux machine, and the whole project will be released
using a GPL~\cite{GNU} license in the near future. 

\section{The entrance test}

The Italian law states that each faculty has to check the entrance
level of freshmen, even if there is neither closed number admittance nor 
a minimum knowledge level. This is particularly 
 useful for scientific and
engineering courses since in this way we can signal students 
whose preparation is below the minimum standard.

After last two years' experience, we decided to monitor only the
mathematical level, fixed at the minimum common part of all kind of
high-schools. We decided to leave out the physics and chemistry tests
since these topics are not taught in some schools (e.g.\ commercial
schools) and in any case our programs start from the very beginning.

The test was organized in 20 questions on the topics listed in
figure~1.
Each question was presented with four possible answers, only one of
which was correct. The score was computed by assigning 3 points to the
right answers, $-1$ to the wrong ones and 0 to the blank ones. The maximum
score was 60, the minimum $-20$, the average score (by random
guessing) is 0, the standard deviation is 20 points, so the test was 
considered passed for a score of 30 or more. 

\begin{figure}[t]
\centerline{\framebox{\includegraphics[width=\columnwidth]{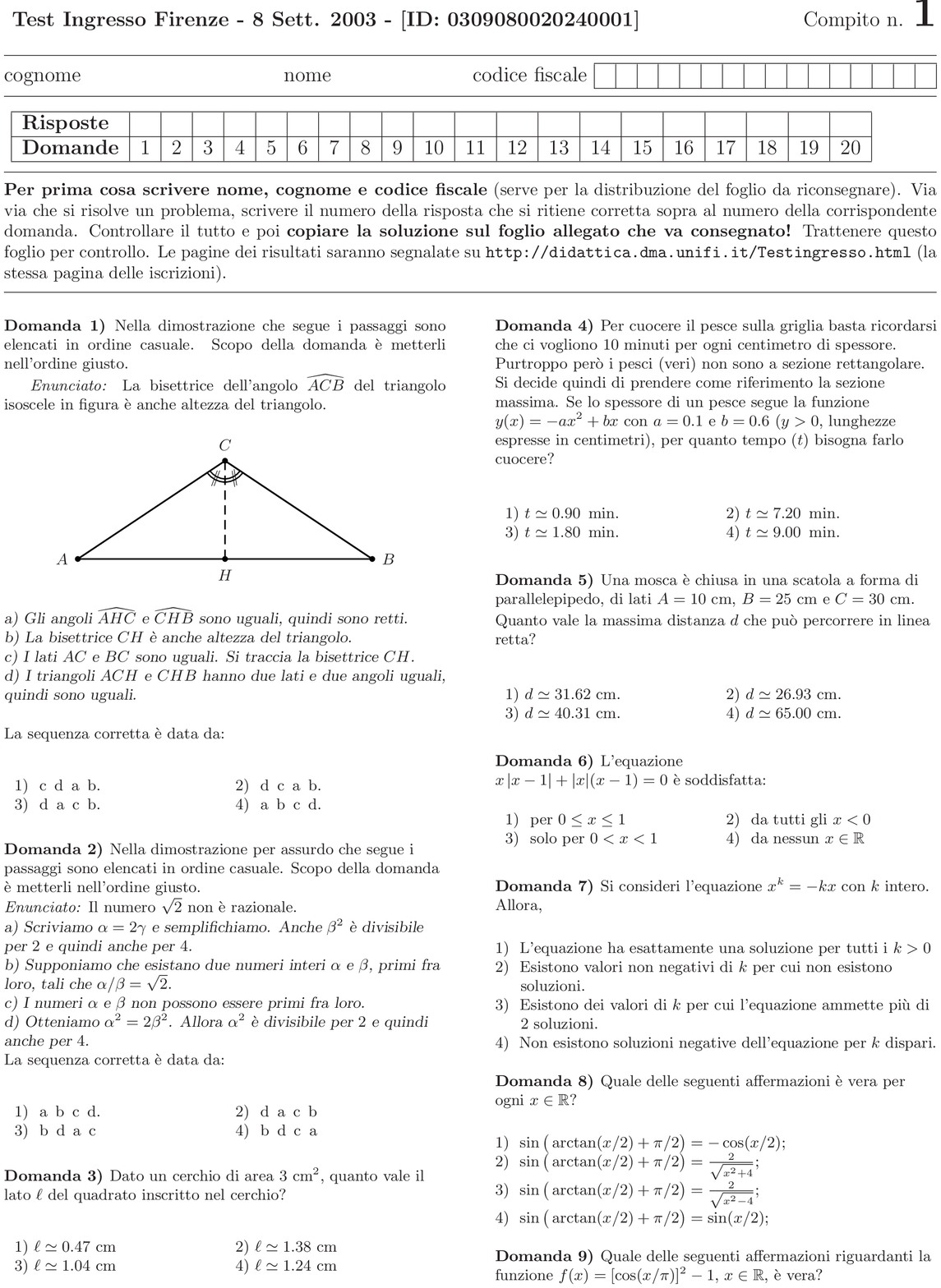}}}
\caption{\label{fig:test} The first page of the test with questions}
\end{figure}

Students had a first test September, $8^{\text{th}}$, then had the
possibility of attending a mathematical refreshment course for two
weeks. Finally there was a supplemental test 
September,  $19^{\text{th}}$. Only
one positive score was sufficient. Those that did not pass any test
are programmed for a third test in November, in the correspondence of
a pause in lecture scheduling. Finally, those that never passed  the
test, are signalled to mathematics teachers that are supposed to
check their base preparation in the correspondence of their first
examination.

\section{Mathematical quiz management with WebTest}

There are many tools that can be used in the preparation of
multiple choice quizzes. However, we established a series of requirements: all
material should be written in \LaTeX (which is the \emph{lingua
franca} of mathematics) and printed using a given template. We also
wanted to have a large number of different but equivalent tests, in
order to discourage copying but avoiding objections. We also wanted to
extract statistics from the answers in order to tune the
mathematical courses. Finally, we wanted to replace the traditional 
handmade checking with an automated one,  but,  
since we decided not to make students pay for
the test, we wanted to avoid specialized hardware for optical reading
and typographic forms. We also knew that students suffer from anxiety
when waiting for a result, even if it is not essential for university
admittance.

Last but not least, we tried to develop a tool simple and powerful
enough that can be also used  for other tasks, like usual examinations
and monitoring comprehension level during courses. 

\begin{figure}[t]
\centerline{\framebox{\includegraphics[width=\columnwidth]{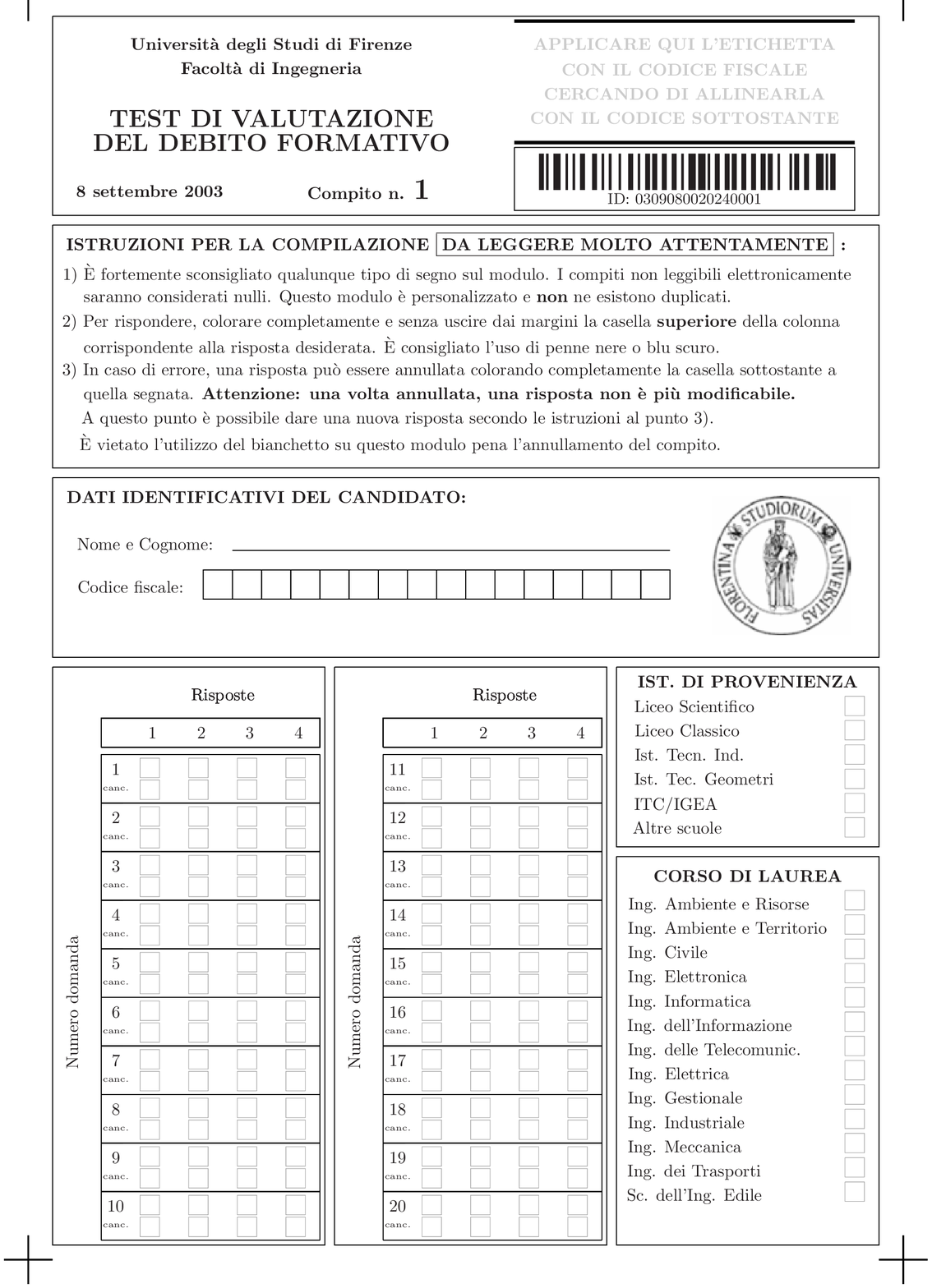}}}
\caption{\label{fig:form} The form for optical reading}
\end{figure}

Traditionally, teachers were asked to prepare a large amount of
questions with answers, that were organized into homogeneous groups,
one for each topic. Then a software samples the questions. However,
this procedure presents two drawbacks: first of all it is difficult to
check all questions and all answers for typographic and logic errors,
and it is also difficult to check for the homogeneity of questions. As
a consequence, a tested set of questions was so a valuable resource
that teachers always tried to recycle them in more than one test. Therefore very repressive methods had to be used to prevent
students from keeping or copying the exercises, and, obviously, there
was a black market of them among students. 

We decided to develop a template system able to variate a given
questions exploiting the combinatorial explosion. Given a certain
number of parameters, the system is able to generate all their
combinations and apply them to the template. This is done in two ways.
For ``textual'' question, one has a given set of right and wrong
answers. In this case we simply sample one from the right set and three from
the wrong one. For numerical questions we have developed a real
parametric template system, able to generate the answer from the 
numerical data, check for unwanted coincidence of answers, for the
correct  range of parameters, etc.

\begin{figure}[t]
\begin{Verbatim}[frame=single,fontsize=\scriptsize]
[% PP={
  R => {descr=>'raggio del cilindro',
   cond=>['R > 0']},
  H => {descr=>'altezza del cilindro', 
   cond=>['H > 0']} %]

[% QQ = BLOCK -%]
Una mosca è chiusa in un bidone cilindrico, 
di raggio di base $R=[%R%]~\centi\metre$   
e altezza  $H=[%H%]~\centi\metre$. 

Quanto vale la massima distanza $d$
che può percorrere in linea retta? 
[%END%] 

[% d = Lcalc.def('\sqrt{4 R^2 + H^2}') %]

[% SS = BLOCK %]
La distanza percorsa dalla mosca è 
l'ipotenusa di un triangolo rettangolo 
di base $2R$ e altezza $H$, quindi 
[% d = Lcalc.def('\sqrt{4 R^2 + H^2}') %]
\[
   d = [%d.s%]~\centi\metre
\]
[% END %]

[% ans = Lcalc.f({
    Right => ['d'],
    Wrong => ['\sqrt{R^2 + H^2}', 'R', 
     '2R', '0.35d', '0.45d'],
  }, '$d\simeq%5.2f~\centi\metre$.');
	
  NewQuestion({Statement=>QQ,Solution=>SS, 
  Answers=>ans,Comment=>CC,Parameters=>PP})%]
\end{Verbatim}
\caption{Source of question 7 of the test in Figure~2.}
\end{figure}

\WebTest is based on the \texttt{TT2}~\cite{TT2}  
Template Toolkit, which already
has a rich set of operators but is quite bad in numerical computation.
So we developed a math plug-in, and after some experiments we decided
that the best thing was to allow teachers to write the parametric
parts using \LaTeX syntax. 

This has two benefits: first of all, not
all mathematics teachers know a computer language, so it is quite
unnatural for them to use an asterisk for multiplication, a double
equal sign for testing equalities, etc. Even most important is the
treatment of variables: in mathematics one is accustomed to
one-letter symbols, using a large font set which includes Greek and
other symbols, which are generally addressed like \verb|\alpha|,
\verb|\beta|, etc. Sometimes one has double-letter symbols, like 
\verb|\Delta S|. And one is also accustomed to using subscript or superscript
for related variables, like $a_1, a^{(2)}$ (written as \verb|a_1|,
\verb|a^{(2)}$|).

\begin{figure}[t]
\begin{Verbatim}[frame=single,fontsize=\scriptsize]
\begin{Problem}{Mosca}
 \Parameter{R}{raggio del cilindro}
 \Cond='R > 0'
 \Parameter{H}{altezza del cilindro}
 \Cond='H > 0'

 \begin{Question}
  \begin{Ask}
   Una mosca è chiusa in un bidone cilindrico, 
   di raggio di base $R=\Val{R}~\centi\metre$   
   e altezza  $H=\Val{H}~\centi\metre$.

   Quanto vale la massima distanza $d$
   che può percorrere in linea retta?
  \end{Ask}

  \Def{d}='\sqrt{4 R^2 + H^2}'

  \begin{Solution}
   La distanza percorsa dalla mosca è 
   l'ipotenusa di un triangolo rettangolo 
   di base $2R$ e altezza $H$, quindi
   \[
      d = \Expr{d}\simeq\FVal{d}~\centi\metre
   \]
  \end{Solution}

  \begin{Answers}
   \Format='$d\simeq%5.2f~\centi\metre$.'
   \Right='d'     \Wrong='\sqrt{R^2 + H^2}' 
   \Wrong='R'     \Wrong='2R'               
   \Wrong='0.35d' \Wrong='0.45d'
  \end{Answers}
 \end{Question}
\end{Problem}
\end{Verbatim}
\caption{Same source as in figure~3 using the \LaTeX syntax.}
\end{figure}

The second benefit is for checking the correctness of expressions. By
using the same syntax for typing a formula and evaluating it, one can
easily write down the formal solution, check it using symbols, and
then print its actual numerical value by exploiting the same expression. 
What happened with the double (\texttt{perl}+\LaTeX) syntax is that one
sees a ``correct'' symbolic expression during checking, but he is not sure that
the same expression is used for numerical computations. 

Using our approach variables are objects, defined using a
\LaTeX expression in a \texttt{TT2} macro. 
A parser converts the expression in
\texttt{Perl},  and checks for  the correctness of the resulting
expression. The actual numerical value of variables is computed at
print time, so that a variation of one parameter immediately propagates
to derived variables like in a computer algebra system.
Variables also carry a format (using \texttt{printf}
syntax) that is used to print their values. In this way one can
perform computations using the full precision, and print the results
with the desired formatting. 

A similar mechanism is used to test
answers for accidental equalities. Generally the ``wrong'' answers are
generated by ``wrong'' formulas that may accidentally coincide
numerically among them or with the right answer. But this coincidence
has to be tested on the printed value, since two numbers may differ
internally yet appear identical when printed. 
So the right and wrong answers are generated using the same format and
only then checked for equality.

The numerical questions are generated by combining several parameters,
and in the definition of the questions one has the possibility of
establishing a set of conditions to be satisfied by parameters and derived
quantities. In this way one can safely use arbitrary parameters, still
avoiding physically incoherent statements (like a contained sphere larger than
the enclosing cube, and so on).

When the 
 template questions are ready, they are expanded using
\texttt{TT2}
and a static database is generated. Teachers that prefer the old method
for generating questions may skip this first part. The database is
then converted into an internal representation which can be dumped in
\texttt{Perl} or \texttt{XML} syntax. 
The actual test structure is specified in \texttt{TT2} (or again \texttt{Perl}
or \texttt{XML}) language, by saying how many questions has to be
sampled by which group, with how many right and wrong answers, if
question order has to be scrambled, and so on. The system then generates
the necessary permutations, and stores them as pointers to the
original database entities.  This data structure
may be saved for faster reload, and is interpolated into a template to
generate a \LaTeX file which is then processed to obtain a \texttt{PDF} file. 

When producing a test, \WebTest also emits a data structure that 
maps each question to the corresponding group (in case of question
scrambling) and associates the right answer. This structure can be
fed to \WebCheck so that students may self-check their answers,
teacher can massively compute the scores from the scanned data, 
and question statistics can be elaborated. 
The use of parametric questions makes statistics not
influenced by the difference in similar questions in the same group. 

The use of text files in all phases make easy the embedding of the
system in a web-based environment or in a more usual graphical GUI. 

This system may be used with no change also to generate quizzes for
every markup system which uses textual representations: \texttt{XML},
\texttt{HTML} and also \texttt{MS-Word} by using \texttt{RTF} files.

The \texttt{TT2} template system uses a quite different syntax respect to
\LaTeX. So we developed a \LaTeX style that allows to use this
language also in the parametric part. This style is polymorphic: used
in the \WebTest environment, it produces the usual \texttt{TT2} source
file, but it may also be used ``at home''  with \LaTeX only and in this
case it simply checks for the correctness of the syntax (using a state
machine) and nicely prints the database for checkout. 

\WebTest may be difficult to install, especially by a
mathematics teacher with no knowledge of computer programming,
expecially if he does not use a \texttt{Unix} environment.
So we developed a \WebWrite plug-in (see section~\ref{sec:WebWrite})
that allows the inclusion of code snippets in web pages with
the display of their output. 
This plug-in, other than offering teachers a web interface
to \WebTest, can be used for teaching computer programming, 
or generally computer-related material like dynamical systems.
The included code runs in a ``jailed'' environment with well-defined
resource limits.

Finally, we adapted the package \texttt{AutoTest}~\cite{AutoTest}
by M. Cesati in order to
optically read the forms filled by students using a generic scanner. 
Our version is able to read the bar-codes that identifies the test and
the students, and the marks of the responses. The input is just a
generic scan, and we were able to use a high-velocity \emph{Xerox} photocopier 
with scanner and network connections  for this task, at no cost (since
it was already present in the University). We are at present extending
the system to be able to use 
fax machines and other scanners as input devices.

\section{WebCheck: Database interface to test management}

The management of examination scheduling is offered by \WebCheck,
which is a set of Perl programs for database interrogations and data
presentation. 

From a student's point of view the first thing to be done is
registration. In the University of Florence, each student is assigned
a unique  id number, which is used as the identifying id in our
database. All Florentine students are already registered in a central
database, which is accessible by themselves through a web interface
for filling-in their study plan and performing other administrative
tasks. We have simply written a web client that, during registration,
tries to connect to this central server using the student's
credentials. If successful, the student is granted access and, if it
is the first time that he/she uses our service, a record is added in
the \WebTeach database. 

However, admittance test is performed by non-registered students, 
and we discovered that freshmen' data  are
entered in the central database with a certain delay, so that the new
students cannot register for their first examinations.

We thus opted for an hybrid scheme: users are allowed also to use
their fiscal code (a unique identifier given by fiscal administrations
to all Italian and foreign people and compulsory for each financial
transaction), and are authenticated either by teachers or by
examination, according to the following scheme. 

Students are asked to register for an examination using the \WebCheck
interface. Their data can be interpolated into a template (using
\texttt{TT2} and \LaTeX) to produce stickers with
identifying bar-codes for each student registered for an examination.
In the meantime, \WebTest produces the tests printout, formed by a
sheet of questions, and a data entry form. This latter form contains a
bar code identifying the test (identifiers again given by \WebCheck).

At test time, teachers distribute the question and the form sheets,
which up to now are anonymous. While students are answering the test,
teachers have plenty of time to check the student's identity and to stick
the corresponding bar-code on the answers' form. 

Teachers use network-enabled photocopier/scanners,
scanners with automatic feeder or even fax machines for digitalizing
the response forms, and then \WebTest can read them and produce a
data file in the right format for \WebCheck. We are at present
making this machinery work automatically, since the test bar-code
contains enough information to address data to the right place in
\WebCheck.  

However, as soon as the test ends, students can connect to \WebCheck and
self-compute their score. We are at present developing an e-mail and
SMS interface, to let students use their mobile phone for the same
task. 

This self-data-entry may also make the optical recognition
obsolete: it is sufficient for teachers to check for the correctness
of these data against the signed form at the moment of the vote
validation (students and teachers must sign an official form).

\section{The WebWrite glueing interface and documentation tool}
\label{sec:WebWrite}
Since a large fraction of engineers exams are dealt using the
\WebCheck system, and students are accustomed to this interface, we
have extended it with a Wiki interface for the distribution of
teaching material. At present, this interface is becoming a generic
tool for web interfaces, and it will include \WebCheck and \WebTest as
plug-ins.

WikiWikiWebs appear as usual web sites in which all pages can be directly 
edited by users through the web browser itself. A simple syntax is used to 
add contents to the web without knowing HTML: the philosophical approach
is to input simple text as one does when writing an e-mail. It is
server's task to present this text in a nice way.

In the Wiki jargon a web page is termed ``topic'', and a homogeneous
set of topics is termed a ``web''. A topic name is usually
distinguished because it is written with uppercase letter in the
middle, as for instance \texttt{WebTeach}.  The presence of a topic
name is automatically recognized,  and the systems adds either the
hyperlink to the corresponding page, or signals the possibility of
creating the missing page.  This favors a top-down approach to
writing: authors start from the index, and then populate the web by
clicking on the orphan links. 

There are several different implementations of WikiWikiWebs, and also
other similar approaches to cooperative environments, either free or
proprietary. We have chosen \texttt{TWiki}~\cite{twiki}, due to its active
community of developers and because it can be extended with
server-side plug-ins (Perl modules). 

Topic contents are stored as text files, allowing the use of normal
UNIX tools like grep to perform searches, RCS for version control and
so on. In the source file the formatting elements are  kept at
minimum: emphasized text is simply surrounded by asterisks or
underscores (an e-mail  convention), bullet lists are marked by
white-spaces followed by an asterisk and a space, URLs are just
plainly written, and so on. Authors are allowed (but discouraged)
to use HTML formatting. 

During the  visualization phase, the text is elaborated in order to
format it as HTML, inserting bold, italics, bullet lists, hyperlinks,
etc. There is the possibility of inserting ``dynamic'' commands to
include other topic, insert the user's name or the date of the day,
and so on. In particular, web indexes are plain pages containing just
a  dynamic command. The text is then embedded into a template,
which furnishes the appropriate ``skin'' including 
buttons for navigation, searches, editing, etc. The
templates are just text files with several dynamic commands.   

The actual template can be selected by site preferences, web
preferences, user preferences or specifying a field in the URL. As
most of \texttt{TWiki} configuration, these preferences are selected by
editing particular pages.  We exploited the template mechanism to
translate the interfaces to Italian, without (almost) affecting the
\texttt{TWiki} code. 

When editing, a simple text area with the source is presented, so
that the author is not distracted by formatting tags. This favors
focusing on contents rather than on appearance.
All topics can have files attached, i.e. uploaded to the server.
This makes simple the distribution of didactic material.
The uploaded files can be linked in the topic text, thus allowing the
inclusion of images/multimedia files in the page shown.

\texttt{TWiki} provides an automatic notification of changes
that can replace bulletin board systems and even mailing lists.
The management of the mailing list is performed by users 
themselves, just by editing a particular page in a web.

\texttt{TWiki} allows the definition of access rights at the level of site,
web-wide and at single page level. User administration
is easy, because the definition of groups of users is itself stored
as a topic (editable only by the administrator group). A hierarchy of
groups can be designed by group-inclusion. The \texttt{TWiki}
authorization mechanism relies on the server ``Basic'' authentication
scheme. We enhanced this mechanism by writing 
a custom \texttt{Apache} handler based on cookies
and interfacing the \WebCheck database, in order to integrate the two systems
without even touching the \texttt{TWiki} sources. 

A plug-in has been written to transparently store in the \WebCheck
database the access-control rules defined in \texttt{TWiki}.
We have replaced the authentication method of \texttt{TWiki} with an Apache
connection handlers, so that we can use the same database for
\WebCheck and \WebWrite access rights. We have also made possible to see
attachments as \texttt{DAV} files, so that they can be edited over a network
connection using a \texttt{DAV}-enabled tool (like \texttt{MS-Office}) 
using the same right access rules of
\texttt{TWiki}. 

There have been several questions from students to be answered about
quizzes. And most of them concerned mathematics. 
We used the \texttt{comment} plug-in of \texttt{TWiki} to set up quickly a rough forum,
and the \texttt{MathMode} one to allow people to use \LaTeX syntax in
questions. In this way we were able to collect answered FAQ from students
without any effort. In the future, we are planning a plug-in to handle
more structured forms of threaded discussions.

\section{Results}

We received a total of  1247 subscriptions for the two tests, for a
total of 957 distinct persons, 
290 of which subscribed to both. 
119 persons subscribed but did not participate
to any test.	The effective number of participants to the tests (i.e.\
that filled the form)
was 540 to the first and 521 to the second one. 

It is interesting to notice that 
317 only participated to the first test, that was sufficient for 
283/540 (52\%).  298 participated to only the second one and 
223 to both,  and 
216 of them did not pass the first test. 
Finally, 120/521 (23\%)  persons passed the second test. 
 
Since in between the two tests we gave a two-week refreshment, it is
interesting to study what happened to the 
223 that participated to both tests. 
3 of them passed both, not decreasing the score. 
4 passed the first and not the second, 
64 did not pass the first and passed the second, and finally 
153 did not pass any test. 

For what concerns the scores (sufficient and not sufficient), 
120 did not decrease and  
103 decreased their score. 
It has to be noticed that, due to
errors in the formulation of problems, two answers were granted right
to everybody for the first test.

\begin{figure}[h]
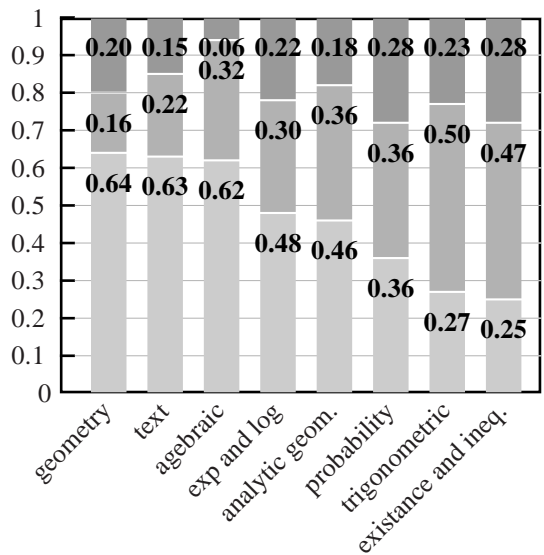

\begin{center}
\renewcommand{\betweenticks}{0.1}
\bardiagrambegin{12}{1}{2cm}{1}{1.5}{0.5cm}{5cm}
\renewcommand{\barlabelangle}{45}
\drawlevellines

\baritem{geometry}{0.64}{green}
\subtopbaritem{}{0.16}{blue}
\subtopbaritem{}{0.20}{red}

\baritem{text}{0.63}{green}
\subtopbaritem{}{0.22}{blue}
\subtopbaritem{}{0.15}{red}

\baritem{agebraic}{0.62}{green}
\subtopbaritem{}{0.32}{blue}
\subtopbaritem{}{0.06}{red}

\baritem{exp and log}{0.48}{green}
\subtopbaritem{}{0.30}{blue}
\subtopbaritem{}{0.22}{red}

\baritem{analytic geom.}{0.46}{green}
\subtopbaritem{}{0.36}{blue}
\subtopbaritem{}{0.18}{red}

\baritem{probability}{0.36}{green}
\subtopbaritem{}{0.36}{blue}
\subtopbaritem{}{0.28}{red}

\baritem{trigonometric}{0.27}{green}
\subtopbaritem{}{0.50}{blue}
\subtopbaritem{}{0.23}{red}

\baritem{existance and ineq.}{0.25}{green}
\subtopbaritem{}{0.47}{blue}
\subtopbaritem{}{0.28}{red}

\bardiagramend{}{}
\end{center}
\caption{\label{fig:statistics} Statistics of answers to questions
grouped by topic. The bottom/middle/top bars represent the percentage
of right/blank/wrong answers to each group.}
\end{figure}

Finally, in figure~\ref{fig:statistics} we report the average
percententage of answers to different kind of questions, which can be
quite useful in tuning the first mathematical courses. 

\section{Conclusions and perspectives}

There are many points we wish to improve in the \WebTeach project,
expecially concerning the integration of the different tools. 

In particular, we would like to develop on-line quizzes in pdf using
the \texttt{AcroTeX}~\cite{Acrotex} package, in a way that makes it
possible to collect statistics for course monitoring. 

All the described software is avaliable upon request, write to authors
for details.

\section*{Acknowledgements}
We wish to thanks all colleagues that elaborated questions and answers, and 
helped in the organization of the test. But most of our thanks are 
devoted to students,  that participated and suggestedi
mprovements to the user interface. The test organizing commitee was 
formed by
Mario Landucci, Giuseppe Anichini, Giovanni Frosali, Marco Spadini,
F.B. and F. M.. This work was supported by the Engineering Faculty
of the University of Florence.

\end{document}